\documentclass[aps,pra,reprint,superscriptaddress]{revtex4-2}
\usepackage{graphicx}
\usepackage{amsmath}
\usepackage{amssymb}
\usepackage{bm}
\usepackage{physics}
\usepackage{subfigure}
\usepackage[colorlinks=true,linkcolor=blue,urlcolor=blue,citecolor=blue]{hyperref}
\usepackage{times}
\usepackage{booktabs}
\usepackage{algorithm}
\usepackage{algpseudocode}
\usepackage{algorithmicx}
\usepackage{svg}
\usepackage{tikz}
\usepackage{listings}
\usepackage{xcolor}
\usepackage{courier} 
\definecolor{codegreen}{rgb}{0,0.6,0}
\definecolor{codegray}{rgb}{0.5,0.5,0.5}
\definecolor{codepurple}{rgb}{0.58,0,0.82}
\definecolor{codeblue}{rgb}{0.25,0.5,0.75}
\definecolor{backcolour}{rgb}{0.96,0.96,0.96}
\definecolor{codeorange}{rgb}{1,0.65,0}

\lstdefinelanguage{Julia}{
    morekeywords={
        abstract,break,case,catch,const,continue,do,else,elseif,
        end,export,false,for,function,immutable,import,importall,if,in,
        macro,module,otherwise,quote,return,switch,true,try,type,typealias,
        using,while
    },
    sensitive=true, 
    morecomment=[l]\#, 
    morecomment=[n]{\#=}{=\#}, 
    morestring=[s]{"}{"}, 
}[keywords,comments,strings]

\begin{document}

\newcommand{\Hop}{\hat{H}}
\newcommand{\im}{{\rm i}}
\newcommand{\rhoop}{\hat{\rho}}
\newcommand{\circuit}{\mathcal{C}}
\newcommand{\supercircuit}{\tilde{\mathcal{C}}}

\newcommand{\EqDef}{\stackrel{\mathrm{def}}{=}}

\newcommand{\gcc}[1]{{\color{red}#1}}

\title{JuliVQC: an Efficient Variational Quantum Circuit Simulator for Near-Term Quantum Algorithms}

\author{Wei-You Liao}
\thanks{These two authors contributed equally}
\affiliation{Henan Key Laboratory of Quantum Information and Cryptography, Zhengzhou,
Henan 450000, China}

\author{Xiang Wang}
\thanks{These two authors contributed equally}
\affiliation{Henan Key Laboratory of Quantum Information and Cryptography, Zhengzhou,
Henan 450000, China}

\author{Xiao-Yue Xu}
\author{Chen Ding}
\author{Shuo Zhang}
\author{He-Liang Huang}
\email{quanhhl@ustc.edu.cn}
\affiliation{Henan Key Laboratory of Quantum Information and Cryptography, Zhengzhou, Henan 450000, China}

\author{Chu Guo}
\email{guochu604b@gmail.com}
\affiliation{Henan Key Laboratory of Quantum Information and Cryptography, Zhengzhou,
Henan 450000, China}


\begin{abstract}
We introduce JuliVQC: a light-weight, yet extremely efficient variational quantum circuit simulator. JuliVQC is part of an effort for classical simulation of the \textit{Zuchongzhi} quantum processors, where it is extensively used to characterize the circuit noises, as a building block in the Schr$\ddot{\text{o}}$dinger-Feynman algorithm for classical verification and performance benchmarking, and for variational optimization of the Fsim gate parameters.
The design principle of JuliVQC is three-fold: (1) Transparent implementation of its core algorithms, realized by using the high-performance script language Julia; (2) Efficiency is the focus, with a cache-friendly implementation of each elementary operations and support for shared-memory parallelization; (3) Native support of automatic differentiation for both the noiseless and noisy quantum circuits. We perform extensive numerical experiments on JuliVQC in different application scenarios, including quantum circuits, variational quantum circuits and their noisy counterparts, which show that its performance is among the top of the popular alternatives. 
\end{abstract}
\maketitle

\section{Introduction}
\begin{figure}[t]
  \centering
  \includegraphics[width=0.85\linewidth]{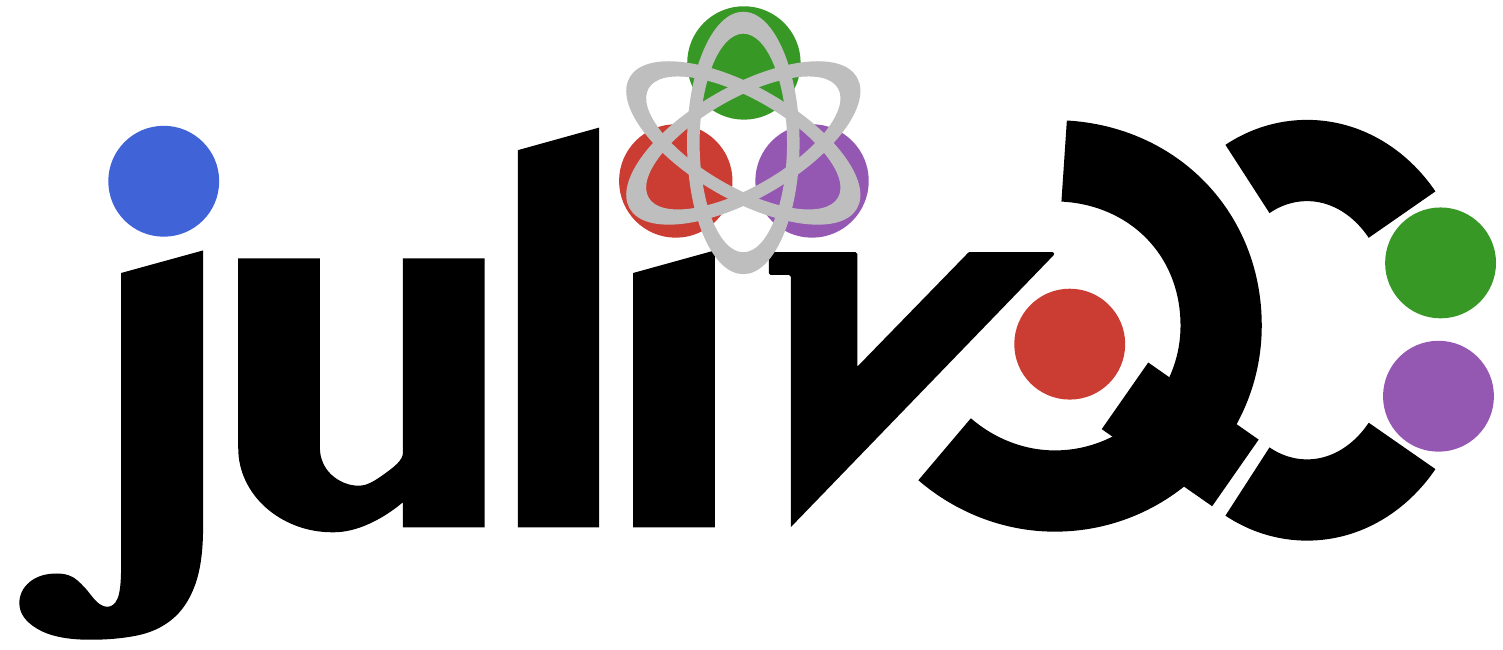}
  \caption{The logo of JuliVQC, which is designed to be a light-weight, yet efficient variational quantum circuit simulator.}
  \label{fig:logo}
\end{figure}

Quantum computing has made enomorous experimental progresses in recent years~\cite{huang2023near,huang2020superconducting}, most prominently the demonstrations of quantum computational advantages~\cite{AruteMartinisQuantumSupremacy2019,WuPan2021,ZhuPan2021,ZhongPan2020,ZhongPan2021,DengPan2023,DengPan2023b,MadsenQuesada2022} and of quantum error correction~\cite{andersen2020repeated,google2021exponential,marques2022logical,zhao2022realization,krinner2022realizing,google2023suppressing,bluvstein2024logical,erhard2021entangling,ye2023logical,wei2024low}. 
However, there is still a considerable level of noise rates in all the existing quantum computing hardware, and there is still a long way before achieving full fault-tolerant quantum computing.

In the noisy intermediate scale quantum (NISQ) computing stage~\cite{preskill2012quantum}, classical simulation of quantum computing is a vital ingredient in developing noisy quantum computers. The applications of classical simulators include: (1) characterizing the noises of quantum computers; (2) helping researchers to explore heuristic quantum algorithms, such as the variational quantum eigensolver~\cite{guo2024experimental,kandala2017hardware,google2020hartree}, without resorting to real quantum computers; (3) benchmarking the performance of quantum computers, such as in the random quantum circuit (RQC) sampling experiments; (4) calibration and optimization of the gate parameters. For application scenario (3), one often needs to simulate relatively shallow quantum circuits but with a large number of qubits (more than $50$), for such task the current state of the art classical simulation algorithm is the tensor network based algorithms~\cite{GuoWu2019,VillalongaMandra2018,VillalongaMandra2019,GrayKourtis2020,HuangChen2021,GuoHuang2021,PanZhang2021,LiuChen2021,PanZhang2021b,ChenYang2022,LiuLiu2022}. For the rest applications, the state-vector simulator is often an ideal choice, since it is numerically exact, efficient for deep quantum circuits, and easily supports efficient classical back propagation. The major drawback of the state-vector simulator is that its memory cost scales exponentially with the number of qubits, which limits its applications mostly within $36$ qubits (see distributed implementation of the state-vector simulator which pushes to $45$ qubits~\cite{HanerSteiger2017}). Nevertheless, for most applications, this issue is either not important, or can be circumvented (such as the Schr$\ddot{\text{o}}$dinger-Feynman simulator which uses the state-vector simulator as the building block, but overcomes the memory issue~\cite{markov2018quantum}). The matrix product state (MPS) based simulator could also be used as a general-purpose simulator, which has been applied to simulate high-depth variational quantum chemistry problems~\cite{ShangLi2022,GuoShang2023}. The efficiency of MPS simulator is mainly limited by the entanglement of the underlying quantum state. In particular, MPS simulators are suited for area-law quantum states with bounded bipartition entanglement, while for volume-law quantum states (which are generally encountered in quantum computing), they will generally not be as efficient as the state-vector simulator.

A plethora of classical simulators have been developed up to date. Popular open-source software include ProjectQ~\cite{ProjectQ}, Cirq~\cite{Cirq}, Qulacs~\cite{Qulacs}, PennyLane~\cite{PennyLane}, Qiskit~\cite{Qiskit}, and QuEST~\cite{QuEST}, which all implement the state-vector simulator (Qiskit also supports the MPS simulator). Those packages provide an end-to-end solution to simulate quantum algorithms. In contrast, the goal of JuliVQC~\cite{JuliVQC} is very simple: JuliVQC only aims to simulate (noisy and variational) quantum circuits in a transparent and  classically the most efficient way. For this purpose, JuliVQC is purely written in the high-performance script language Julia, to enable code transparency and efficiency at the same time. It is also implemented in a cache-friendly way and fully supports shared-memory parallelization to maximize efficiency. Moreover, classical automatic differentiation are fully supported for both the noiseless and noisy quantum circuits which exploits the reversibility of (noisy) quantum circuits to minimize the memory usage and computational cost simultaneously. On the other hand, only a minimal effort is made in JuliVQC to manipulate quantum circuits. No effort is made to support the high-level operations such as quantum compilation. Therefore, JuliVQC may also be used as a backend for existing classical simulation frameworks for quantum computing. JuliVQC has been extensively used in the development of the \textit{Zuchongzhi} quantum processors~\cite{WuPan2021,ZhuPan2021}, for at least three scenarios: (i) it is used to verify quantum circuits with less than $36$ qubits; (ii) it is used as the building block in the Schr$\ddot{\text{o}}$dinger-Feynman simulator for verification and performance benchmarking of quantum circuits with more than $36$ qubits; (iii) it is used as a variational quantum circuit simulator to optimize the Fsim gate parameters.

The paper is organized as follows: in Sec.~\ref{sec:usage}, we show the pipeline of running (variational) quantum circuits simulations using JuliVQC. In Sec.~\ref{sec:algs}, we show the major implementation-wise techniques used in JuliVQC, including the cache-friendly implementation of the elementary gate operations on the state vector and their parallelization, and the algorithms for automatic differentiation of (noisy) quantum circuits. In Sec.~\ref{sec:Performance}, we benchmark the performance of JuliVQC against some popular open source quantum circuit simulators. We summarize and outlook in Sec.~\ref{sec:summary}.

\section{Overview of JuliVQC}\label{sec:usage}
In this section we show the working pipeline of JuliVQC for simulating quantum circuits and variational quantum circuits, as well as their noisy counterparts.

\subsection{State initialization}
The first step of using JuliVQC for any quantum circuit simulation is to initialize a quantum state stored as a state vector. JuliVQC provides twos function: \texttt{StateVector} and \texttt{DensityMatrix} to initialize a pure state and a mixed state respectively. For \texttt{StateVector}, one could either provide an integer $n$ as input which initialize an $n$-qubit pure quantum state with each qubit in state $\vert 0\rangle$, or a one-dimensional array of size $2^n$ as input which will directly be used as the data of the pure state. For \texttt{DensityMatrix}, one could provide an integer $n$ as input which initialize an $n$-qubit mixed quantum state with each qubit in state $\vert 0\rangle\langle 0\vert$, one can also provide a one-dimensional array of size $2^{2n}$, or equivalently a two-dimensional array of size $2^n\times 2^n$ as input which will directly be used as the data (one need to make sure that the array is a proper density matrix, e.g., semi-positive with unit trace). One could also provide an optional type parameter as the first argument of these two functions to enforce the scalar type used for the quantum state, such as \texttt{StateVector(Float64, n)}. The illustrative code for initializing two-qubit pure and mixed quantum states is shown in Listing 1.



Mathematically, the data of an $n$-qubit pure state should be understood as a rank-$n$ tensor, and the data of an $n$-qubit mixed state should be understood as a rank-$2n$ tensor, where each dimension has size $2$. 
As implementation-wise details, the qubits are internally labeled from $1$ to $n$ for pure state , while for mixed state the ket indices are labeled from $1$ to $n$ and the bra indices are labeled from $n+1$ to $2n$. 
Column-major storage is used for the data of both pure and mixed quantum states, e.g., the smaller indices of the tensor are iterated first. These details are not important for the users if they do not want to access the raw data of the quantum states.

\begin{lstlisting}[caption={Initializing pure and mixed quantum states}, label=lst:listing1]
  using JuliVQC

  n = 2
  pure_state = StateVector(n)
  mixed_state = DensityMatrix(n)
  custom_pure_state= StateVector(Float64, [0,1,0,0])
  custom_mixed_state = DensityMatrix([0,0,0,0,0,0,0,0,0,0,0,0,0,0,0,1])
\end{lstlisting}

\subsection{Initializing quantum gates and quantum channels}
The second step of using JuliVQC is to build a quantum circuit, for which one needs to define each elementary quantum gate operations (and quantum channels for noisy quantum circuits). 

The universal way of defining quantum gates is to use the function \texttt{QuantumGate(positions, data)}, where the first argument specifies the qubits indices that the gate operates on, for example $\texttt{positions} =(1, 3)$, and the second argument is the raw data of the gate operation which should be a unitary matrix. 
Again, the raw data provided to \texttt{QuantumGate} should be stored in column-major storage (the matrix should be understood as a rank-$2m$ tensor if it acts on $m$ qubits), which is important for the user to use this function correctly. Since many textbook definitions of gate operations use the row-major storage of the tensor indices, we also provide a utility function \texttt{n\_qubits\_mat\_from\_external}, which accepts the raw data of a quantum gate as a row major matrix and outputs it as a column-major matrix.

In the meantime, JuliVQC provides specialized definitions of commonly used quantum gates, which include the single-qubit operations \texttt{XGate}, \texttt{YGate}, \texttt{ZGate}, \texttt{HGate}, \texttt{SGate}, \texttt{TGate}, \texttt{sqrtXGate} (square root of \texttt{XGate}), \texttt{sqrtYGate} (square root of \texttt{YGate}), the two-qubit gate operations \texttt{SWAPGate}, \texttt{iSWAPGate}, \texttt{CZGate}, \texttt{CNOTGate}, the three-qubit gate operations \texttt{TOFFOLIGate}, \texttt{FREDKINGate}. The general usage of these predefined non-parametric gate operations is \texttt{G(i)} or \texttt{G((i,))} if \texttt{G} a single-qubit gate, \texttt{G(i,j)} or \texttt{G((i,j))} if \texttt{G} a two-qubit gate and \texttt{G(i,j,k)} or \texttt{G((i,j,k))} if \texttt{G} a three-qubit gate, where \texttt{i,j,k} are integers. JuliVQC also provides general two-qubit and three-qubit controlled gate operations: \texttt{CONTROLGate} and \texttt{CONTROLCONTROLGate}, which can be used as \texttt{CONTROLGate(i,j,data)} (\texttt{i} is the control qubit and \texttt{j} is the target qubit) and \texttt{CONTROLCONTROLGate(i,j,k,data)} (\texttt{i} and \texttt{j} are control qubits and \texttt{k} is the target qubit), with \texttt{data} the raw data for the target single-qubit operation. 
Although one could directly build all these predefined gate operations using the general \texttt{QuantumGate} function, specific optimizations have been implemented for most of the predefined gate operations by exploring their structures, which will usually be faster than using the \texttt{QuantumGate} function.

JuliVQC also supports parametric quantum gates as a necessary ingredient for variational quantum circuits. The predefined single-qubit parametric quantum gates include \texttt{RxGate} (rotational \texttt{XGate}), \texttt{RyGate} (rotational \texttt{YGate}), \texttt{RzGate} (rotational \texttt{ZGate}) which are parameterized as
\begin{align}
\texttt{Rx}(\theta) &= \begin{pmatrix} \cos\frac{\theta}{2} & -\im \sin\frac{\theta}{2} \\ -\im \sin\frac{\theta}{2} & \cos\frac{\theta}{2} \end{pmatrix} ; \\
\texttt{Ry}(\theta) &= \begin{pmatrix} \cos\frac{\theta}{2} & - \sin\frac{\theta}{2} \\ - \sin\frac{\theta}{2} & \cos\frac{\theta}{2} \end{pmatrix} ; \\
\texttt{Rx}(\theta) &= \begin{pmatrix} e^{-\im \frac{\theta}{2}} & 0 \\ 0 & e^{\im \frac{\theta}{2}} \end{pmatrix} 
\end{align}
respectively. The two-qubit controlled rotational gates are also supported, including \texttt{CRxGate}, \texttt{CRyGate}, \texttt{CRzGate}, which are controlled \texttt{RxGate}, \texttt{RyGate}, \texttt{RzGate} respectively. These single-qubit and two-qubit gates are parameterized by a single parameter. In addition, JuliVQC also predefines the \texttt{FSIMGate} with $5$ parameters:
\begin{align}
&\texttt{FSIM}(\theta, \phi, \Delta_+, \Delta_-, \Delta_{-, {\rm off}}) = \nonumber \\
 &\begin{pmatrix} 1 & 0 & 0 & 0 \\ 0 & e^{\im (\Delta_+ + \Delta_-)}\cos\theta & -\im e^{\im (\Delta_+ - \Delta_{-, {\rm off}})}\sin\theta & 0 \\ 0 & -\im e^{\im (\Delta_+ + \Delta_{-, {\rm off}})}\sin\theta & e^{\im (\Delta_+ - \Delta_-)}\cos\theta & 0 \\ 0 & 0 & 0 & e^{\im (2\Delta_+ - \phi)} \end{pmatrix}.
\end{align}
The general interface for initializing a parametric quantum gate is \texttt{G(i..., paras; isparas)} where \texttt{paras} is a single scalar if \texttt{G} only has a single parameter or an array of scalars if \texttt{G} has several parameters. The keyword \texttt{isparas} has the same size as \texttt{paras}, namely it could be a single \texttt{Bool} type or an array of \texttt{Bool}, which is used to specify which parameters in \texttt{paras} are really treated as variational parameters. The default value of \texttt{isparas} is false, which means that the gate will actually be treated as a non-parametric gate.

The illustrative code for initializing non-parametric and parametric quantum gates is shown in Listing 2.

\begin{lstlisting}[caption={Initializing quantum gates and parametric quantum gates}, label=lst:listing2]
    using JuliVQC

    n=1
    X = XGate(n)
    ncontrol = 1
    ntarget = 2
    CNOT = CNOTGate(ncontrol, ntarget)
    theta = pi/2
    non_para_Rx = RxGate(n, theta, isparas=false) # a non-parametric Rx gate
    para_Rx = RxGate(n, theta, isparas=true) # a parametric Rx gate
\end{lstlisting}


In additional to the quantum gate operations, an indispensable ingredient for noisy quantum circuit is the quantum channel, which describes the effects of noises. 
Mathematically, a quantum channel $\Lambda$ is a positive semi-definite and trace-preserving map on the mixed quantum state $\rhoop$, which can be generally written as 
\begin{align}\label{eq:channel}
  \Lambda(\rhoop)=\sum_iK_i\rhoop K_i^\dagger,
\end{align}
where $K_i$s are the Kraus operators satisfying the completeness relation $\sum_iK_i^\dagger K_i=I$, with $I$ the identity matrix. 
Similar to the function \texttt{QuantumGate}, JuliVQC provides a universal function \texttt{QuantumMap(positions, kraus)} which allows the user to define arbitrary quantum channels, where the first argument \texttt{positions} specifies the qubit indices that the quantum channel operates on, similar to the case of a unitary quantum gate, and the second argument \texttt{kraus} is a list of Kraus operators.
JuliVQC also provides some commonly used single-qubit quantum channels based on the function \texttt{QuantumMap}, including \texttt{AmplitudeDamping(pos; $\gamma$)} with
\begin{align}
  \begin{gathered}
    K_{0} =\begin{pmatrix}1&0\\0&\sqrt{1-\gamma}\end{pmatrix}, 
    K_{1} \left.=\left(\begin{matrix}0&\sqrt{\gamma}\\0&0\end{matrix}\right.\right), 
  \end{gathered}
\end{align}
\texttt{PhaseDamping(pos; $\gamma$)} with 
\begin{align}
  \begin{gathered}
    K_{0} =\begin{pmatrix}1&0\\0&\sqrt{1-\gamma}\end{pmatrix}, 
    K_{1} \left.=\left(\begin{matrix}0&0\\0&\sqrt{\gamma}\end{matrix}\right.\right), 
  \end{gathered}
\end{align}
and \texttt{Depolarizing(pos; p)} with 
\begin{align}
  \begin{gathered}
    K_{0} =\sqrt{1-\frac{3p}{4}}\begin{pmatrix}1&0\\0&1\end{pmatrix}, 
    K_{1} \left.=\sqrt{p/2}\left(\begin{matrix}0&1\\1&0\end{matrix}\right.\right), \\
    K_{2} \left.=\sqrt{p/2}\left(\begin{array}{cc}0&-\im\\ \im&0\end{array}\right.\right), 
    K_{3} =\sqrt{p/2}\begin{pmatrix}1&0\\0&-1\end{pmatrix}.
  \end{gathered}
\end{align}
The first argument \texttt{pos} in these predefined single-qubit quantum channels is an integer specifying the qubit index being operated on, and the keyword arguments $\gamma$ or \texttt{p} are parameters describing the strength of noises. A quantum channel is currently not allowed to contain variational parameters.


\subsection{Manipulating quantum circuits}
JuliVQC uses a very simple wrapper \texttt{QCircuit} on top of an array of quantum operations to represent a quantum circuit.
Each element of \texttt{QCircuit} can be either a (parametric) unitary gate operation, a quantum channel, or a \texttt{QCircuit}. 
Once a \texttt{QCircuit} object, denoted as \texttt{circ}, has been built, one can extract all its variational parameters using the function \texttt{active\_parameters(circ)}, one can also reset the parameters in \texttt{circ} with the function \texttt{reset\_parameters!(circ, paras)}, where \texttt{paras} is an array of scalars used to replace the existing variational parameters. In addition, JuliVQC provides a function \texttt{fuse\_gates(circ)} which performs preliminary simplification of a given \texttt{QCircuit} and can only be used for noiseless quantum circuits currently. The logic behind this function is very simple: if there is a single-qubit gate on $i$ which is right before or after a two-qubit gate on $i$ and $j$, then the single-qubit gate is absorbed into the two-qubit gate. The negative effect of this function is that all the gate operations will be converted into non-parametric gates, and specialized gates will be converted into the most general \texttt{QuantumGate}.
In Listing 3, we show various operations on \texttt{QCircuit}.


\begin{lstlisting}[caption={Building quantum circuits}, label=lst:listing3]
    using JuliVQC

    state=StateVector(2)
    circ = QCircuit([XGate(1), Dephasing(1, p=0.3)) ,RxGate(2, pi/2, isparas = true)]
  \end{lstlisting}

\subsection{Running quantum algorithms}
After initializing the quantum state and building the quantum circuit, one could apply the quantum circuit onto the quantum state using the \texttt{apply!(circ, state)} function (\texttt{state} can either be a pure state or a density matrix), which modifies the quantum state in-place. There is also an out-of-place version of this operation, e.g., \texttt{apply(circ, state)} or equivalently \texttt{circ * state}, which will return a new quantum state and is useful for running variational quantum algorithms. 

A standard quantum algorithm ends by measuring some or all the qubits. JuliVQC provides a function \texttt{measure!(state, i)}, which measures the $i$-th qubit and collapses the quantum state in-place. This function will return a tuple, where the first one is the measurement outcome ($0$ or $1$), and the second one is the exact probability of the measurement outcome (We note that the probability can not be directly obtained from a quantum computer). 
In Listing 4, we show a standard quantum algorithm by applying a quantum circuit onto a quantum state and ends with a quantum measurement.

\begin{lstlisting}[caption={Runing quantum algorithms}, label=lst:listing4]
    using JuliVQC

    state = StateVector(2)
    circuit = QCircuit([HGate(1), RyGate(1,pi/4,isparas = false) ,CNOTGate(1,2)])
    apply!(circuit,state)
    outcome, prob = measure!(state,2)
  \end{lstlisting}

\subsection{Building qubit operators}
An important ingredient for variational quantum algorithms is to compute the expectation value of some qubit operator, which is a summation of Pauli strings. The qubit operator is represented as a \texttt{QubitsOperator} object in JuliVQC, which can be built as in Listing 5. Once a qubit operator \texttt{op} has been initialized, one could apply the function \texttt{expectation(op, state)} to evaluate the expectation of it on the quantum state \texttt{state} (Again this operation can not be directly performed on a quantum computer).
\begin{lstlisting}[caption={Building qubit operators}, label=lst:listing5]
  using JuliVQC

  function heisenberg_1d(L; hz=1, J=1)
      terms = []
      # one site terms
      for i in 1:L
          push!(terms, QubitsTerm(i=>"z", coeff=hz))
      end
      # nearest-neighbour interactions
      for i in 1:L-1
          push!(terms, QubitsTerm(i=>"x", i+1=>"x", coeff=J))
          push!(terms, QubitsTerm(i=>"y", i+1=>"y", coeff=J))
          push!(terms, QubitsTerm(i=>"z", i+1=>"z", coeff=J))
      end
      return QubitsOperator(terms)
  end
\end{lstlisting}

\subsection{Running variational quantum algorithms}
JuliVQC has a transparent support for automatic differentiation, one could simply run a variational quantum algorithm in the similar way as a standard quantum algorithm, as demonstrated in Listing 6. 

\begin{lstlisting}[caption={Running variational quantum algorithms}, label=lst:listing6]
    using JuliVQC, Zygote

    state = StateVector(3)
    op = heisenberg_1d(3) #Construct Heisenberg Hamiltonian as a qubit operator
    alpha = 0.01
    circ = QCircuit()
    for depth in 1:4
        for i in 1:2
            push!(circ,CNOTGate(i,i+1))
        end
        for i in 1:3
            push!(circ,RyGate(i,randn(),isparas=true))
            push!(circ,RxGate(i,randn(),isparas=true))
        end     
    end

    loss(circ)=real(expectation(op, circ * state))
    grad = gradient(loss, circ)[1] # calculate gradient
    paras = active_parameters(circ) # extracting the parameters
    new_paras = paras - alpha * grad # gradient descent to update parameters
    reset_parameters!(circ, new_paras) # reset parameters
  \end{lstlisting}

The major difference from running a standard quantum algorithm is that one wraps the \texttt{expectation} function into a loss function, and then use the function \texttt{gradient(loss, circ)} to obtain the gradient of the parameters within the quantum circuit. 
Under the hood, the gradient is calculated using the Zygote auto-differentiation framework, by rewriting the backpropagation rules of a few elementary operations (the detailed algorithm we use to implement the classical backpropagation will be shown later).

\section{Core algorithms used in JuliVQC}\label{sec:algs}

\subsection{The Schrödinger algorithm for noiseless quantum circuit}


We first give a brief introduction to the Schr$\ddot{\text{o}}$dinger algorithm for simulating noiseless quantum circuits, which is the mathematical building block of JuliVQC.
An $n$-qubit pure quantum state can be generally written as:
\begin{align}
  \label{eq1}
  |\phi\rangle=\sum_{\sigma_1,\sigma_2,...,\sigma_n}c_{\sigma_1,\sigma_2,...,\sigma_n}|\sigma_1,\sigma_2,\ldots,\sigma_n\rangle,
\end{align}
where $|\sigma_1,\sigma_2,\ldots,\sigma_n\rangle$ represents a specific computational basis and $c_{\sigma_1,\sigma_2,...,\sigma_n}$ is the amplitude (which is a complex number in general) of it.
The application of a quantum circuit $\circuit$, which consists of $M$ unitary quantum gate operations denoted as $\hat{Q}_j$ with $1\leq j\leq M$, onto a pure quantum state initialized as $|0^{\otimes n}\rangle $ can be denoted as 
\begin{align}
  \label{eq:circuitevolve}
  |\psi\rangle=\circuit|0^{\otimes n}\rangle=\hat{Q}^M\cdots\hat{Q}^1|0^{\otimes n}\rangle.
\end{align}

In the Schr$\ddot{\text{o}}$dinger algorithm, all the amplitudes (there are $2^n$ in total) are directly stored in memory, and each quantum gate operation updates all (or most of) these amplitudes.
For example, a general single-qubit gate operation on the $i$-th qubit, denoted as a $2\times 2$ matrix $\hat{Q}_{\sigma_{i}}^{\sigma_{i}^{\prime}}$, can be written as
\begin{align}
  \label{eq3}
  c_{\sigma_{1},...,\sigma_{i}^{\prime},...,\sigma_{n}} \leftarrow \sum_{\sigma_{i}}\hat{Q}_{\sigma_{i}}^{\sigma_{i}^{\prime}}c_{\sigma_{1},...,\sigma_{i},...,\sigma_{n}},
\end{align}
a general two-qubit gate operation on the $i$-th and $j$-th qubits, denoted as a $2\times 2\times 2\times 2$ tensor $\hat{Q}_{\sigma_{i},\sigma_{j}}^{\sigma_{i}^{\prime},\sigma_{j}^{\prime}}$, can be written as 
\begin{align}
  \label{eq:twoqubitgate}
  c_{\sigma_{1},\ldots,\sigma_{i}^{\prime},\ldots,\sigma_{j}^{\prime},\ldots,\sigma_{n}}=\sum_{\sigma_{i},\sigma_{j}}\hat{Q}_{\sigma_{i},\sigma_{j}}^{\sigma_{i}^{\prime},\sigma_{j}^{\prime}}c_{\sigma_{1},\ldots,\sigma_{i},\ldots,\sigma_{j},\ldots,\sigma_{n}}.
\end{align}
As an example, we show a simple implementation of a two-qubit gate operation using three for-loops in Algorithm.~\ref{alg:straightforward contraction}, which modifies the state vector in place.

\begin{algorithm}[H]
  \caption{straightforward tensor contraction algorithm}\label{alg:straightforward contraction}
  \begin{algorithmic}[1]
  \State Reshaping: $ \tilde{c}_{(\sigma_1,...,\sigma_{i-1}),\sigma_i,(\sigma_{i+1},...,\sigma_{j-1}),\sigma_j,(\sigma_{j+1},...,\sigma_n)} \leftarrow c_{\sigma_1,\sigma_2, ...,\sigma_n}$
  \For{$s_3=0:2^{n-j}-1$}
    \For{$s_2=0:2^{n-j}-1$}
      \For{$s_1=0:2^{n-j}-1$}
  
      \State $\tilde{c}_{s_1,\sigma_i',s_2,\sigma_j',s_3}=\sum_{\sigma_i,\sigma_j}\hat{Q}_{\sigma_i,\sigma_j}^{\sigma_{i'},\sigma_{j'}}\times \tilde{c}_{s_1,\sigma_i,s_2,\sigma_j,s_3}$
      \EndFor
    \EndFor
  \EndFor

  \end{algorithmic}
  \end{algorithm}

\begin{figure}[t]
  \centering
  \includegraphics[width=\columnwidth]{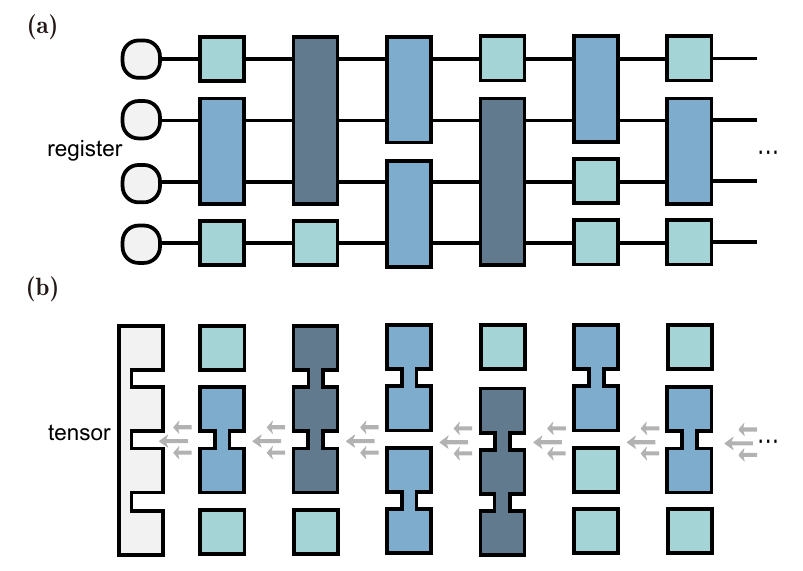}
  \caption{(a) Applying a quantum circuit onto a $4$-qubit quantum state, which is mapped into the contraction of a tensor network in (b). The tensor network contraction in (b) is performed from left to right.}
  \label{fig:tensor_constraction}
\end{figure}

Overall, the Schr$\ddot{\text{o}}$dinger algorithm can also be viewed as the contraction of a tensor network, but with a specific contraction order: the low-rank tensors representing the quantum gate operations are sequentially absorbed into the rank-$n$ tensor representing the pure quantum state, as shown in Fig.~\ref{fig:tensor_constraction}.

\subsection{Cache-friendly implementation of gate operations and shared-memory parallelization}

\begin{figure}[t]
  \centering
  \includegraphics[width=\columnwidth]{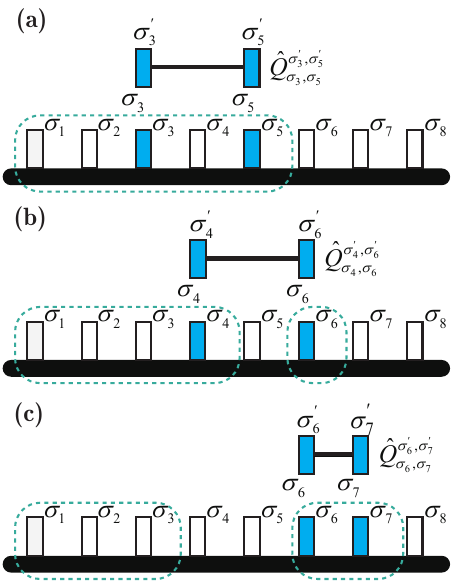}
  \caption{Scheme to aggregate a fixed size matrix for the inner matrix-matrix multiplication of the two-qubit gate operation $\hat{Q}_{\sigma_i,\sigma_j}^{\sigma_{i'},\sigma_{j'}}$ for case (a) both indices $i$ and $j$ been contracted are less than $5$; (b) one of them is less than $5$ and (c) both of them are larger than $5$. The tensor indices inside the blue dashed boxes mean that they should be taken for the inner matrix-matrix multiplication, while the rest indices are iterated over in the outer for-loops.}
  \label{fig:aggregate scheme}
\end{figure}

A general multi-qubit gate operation can be implemented similar to Algorithm.~\ref{alg:straightforward contraction}, where the for-loops iterate over the uncontracted indices, and the inner operation performs essentially a matrix-vector multiplication. However, this approach is highly inefficient on classical computers due to the low compute density. For example, the inner operation in Algorithm.~\ref{alg:straightforward contraction} consists of a $4\times4$ matrix multiplying a four-element vector. Consequently, the total number of floating-point operations (FPOs) is $4\times2^n$. Meanwhile, each element of the state vector must be moved from main memory to the CPU cache and back for at least once, leading to a memory access complexity of $O(2\times2^n)$. Therefore the number of memory accesses is roughly half of the number of FPOs, a situation that is unfriendly to modern computer hardware.

\begin{figure}[t]
  \centering
  \includegraphics[width=1
  \linewidth]{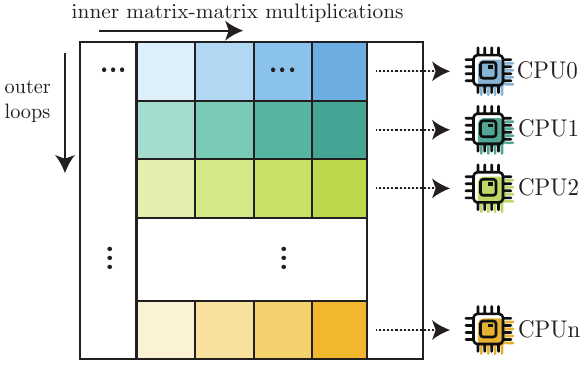}
  \caption{Shared-memory parallelization of each quantum operation. Each square represents the aggregated matrix-matrix multiplication inside the for-loops. Since the matrix-matrix multiplications only accesses non-overlapping data, the outer for-loops can be perfectly parallelized.}
  \label{fig:shared-memory architecture}
\end{figure}

Although it is impossible to improve the compute density of a single gate operation (however, it is possible to increase the overall compute density by fusing several gate operations together~\cite{ZhangZhai2022}), one could still greatly increase the computational efficiency by vectorizing the inner matrix-vector multiplication and making better use of the cache. 
To achieve this, we aggregate $8$ vectors of size $4$ for the inner operation of Algorithm.~\ref{alg:straightforward contraction}, such that it becomes the multiplication of two matrices of sizes $4\times 4$ and $4\times 8$. This can be done in Fig.~\ref{fig:aggregate scheme}(a,b,c), depending on whether the qubit indices $i$ and $j$ are less than $5$ (the optimal value of this number will of course be dependent on the inner most cache size) or not. Importantly, in all the three cases, there are at least $8$ elements that could be taken contiguously, which could thus be vectorized by most modern computing hardware and also make better use of the cache size. 
Furthermore, the for-loops of Algorithm.~\ref{alg:straightforward contraction} can easily be parallelized as the inner operation acts on non-overlapping segments of the state vector, which is illustrated in Fig.~\ref{fig:shared-memory architecture}.


\subsection{Automatic differentiation of noiseless parametric quantum circuits}
\label{subsection:Automatic Differentiation}

Automatic differentiation (AD) is a critical technique for efficiently and accurately computing derivatives by computer programs. AD is implemented in two principal modes: forward mode and reverse mode. The forward mode is best suited for functions with more outputs than inputs, whereas reverse mode is advantageous for functions with more inputs than outputs. Therefore reverse-mode AD is more relevant for optimization problems as the loss function generally outputs a single scalar.

Generally, the idea of reverse-mode AD could be roughly sketched as follows: for a loss function which is composed of many elementary functions, one could represent the computational flow from the input to output as a graph, then as long as one can properly define the ``adjoint'' function of each elementary function which back propagates the gradients, one can traverses the computational graph in reverse order to obtain the gradient of the loss function. 
The advantage of reverse-mode AD is that the computational cost of the backpropagation is roughly the same as evaluating the loss function itself~\cite{autodiff}, while the possible disadvantage is that one needs to store all the intermediate outputs along the computational graph.

For variational quantum circuit simulators which ends by calculating the expectation value of some qubit operators, there are only two elementary functions involved, the quantum gate operation and the \texttt{expectation} function, as the quantum circuit evolution can be viewed as a composed function of many quantum gate operations. Therefore, in principle, one could define the adjoint function these two functions and then rely on any AD framework to calculate the gradient of a variational quantum circuit (see Ref.~\cite{GuoPoletti2021} for general rules of defining the adjoint functions).

However, this approach will be extremely inefficient, since if one treats each quantum gate operation as an elementary function, one needs to store as many copies of the quantum state as the number of quantum gates. To overcome this issue, one could treat the whole circuit evolution function in Eq.(\ref{eq:circuitevolve}) as an elementary function, 
and then recompute all those intermediate quantum states on the fly by reversely evolving the quantum circuit. The advantage of this approach is that only two copies of quantum states need to be stored, with a minor computational overhead that one needs to perform the reverse evolution of the quantum circuit (therefore the computational cost of the backpropagation is roughly two times that of evaluating the loss function). This memory-efficient approach has been thoroughly discussed in Ref.~\cite{GuoShang2023} and implemented for the MPS simulator. Here we implement this algorithm for the state-vector simulator and integrate it into the Zygote AD framework. In the following we briefly sketches it since a similar approach will be used later for implementing the AD for noisy variational quantum circuits.

Consider a quantum state that evolves from $|0^{\otimes\mathrm{n}}\rangle$ to $|\psi(\theta)\rangle$ through a series of parameter gates $\hat{Q}(\theta_M)\cdots\hat{Q}(\theta_2)\hat{Q}(\theta_1) $, the cost function, defined as the expectation value of a qubit operator $\hat{H}$, is denoted as:
\begin{align}\label{eq:loss_pure}
\mathcal{L}(\theta)=\langle0^{\otimes n}|\hat{C}^\dagger(\theta)\hat{H}\hat{C}(\theta)|0^{\otimes n}\rangle .
\end{align}
The derivative of $\mathcal{L}(\theta)$ with respect to a parameter $\theta_j$ is given by:
\begin{align}\label{eq:diff_pure}
  \frac{\partial\mathcal{L}(\theta)}{\partial\theta_j}=\langle\psi(\theta)|\hat{H}\hat{C}_{M:j+1}\frac{\mathrm{d}\hat{Q}(\theta_j)}{\mathrm{d}\theta_j}\hat{C}_{j-1:1}|0^{\otimes n}\rangle+\mathrm{H.c.},
\end{align}
where $\hat{C}_{b:a}\quad=\hat{Q}(\theta_b)\hat{Q}(\theta_{b-1})\cdots\hat{Q}(\theta_a)$.Then by defining $|\Phi_{j}\rangle=\hat{C}_{j:1}|0^{\otimes n}\rangle,$ and $\langle\Psi_{j}|=\langle\psi(\theta)|\hat{H}\hat{C}_{M:j+1}, $ Eq.~\eqref{eq:diff_pure} can be simplified to:
\begin{align}\label{eq:diff_pure2}
\frac{\partial\mathcal{L}(\theta)}{\partial\theta_j}=\langle\Psi_j|\frac{\mathrm{d}\hat{Q}(\theta_j)}{\mathrm{d}\theta_j}|\Phi_{j-1}\rangle+\mathrm{H.c.}.
\end{align}
Therefore, by caching the two intermediate ``quantum states'' $|\Phi_m\rangle $ and $|\Psi_{m}\rangle = \hat{H}|\psi(\theta)\rangle$ (we note that the later is not a proper quantum state since it may not be normalized), one can compute the gradient by using only these two copies. This technique is outlined in Algorithm.\ref{alg:noiseless ad}.

\begin{algorithm}[H]
\caption{Memory-efficient way to evaluate Eq.(\ref{eq:diff_pure}), with two input states $\vert \Phi\rangle$ and $\vert\Psi_M\rangle$.} \label{alg:noiseless ad}
\begin{algorithmic}[1]
\State Initialization: $\vert  \Phi\rangle = \vert \Phi_M \rangle$, $\vert \Psi\rangle=\Hop \vert \Psi_M \rangle$ and $grads = zeros(M)$
\For{$j = M:-1:1$}
    \State $ \vert \Phi \rangle \leftarrow \hat{Q}(\theta_{j})^{-1} \vert \Phi\rangle   $
    \State $grads[j] = 2\real(\langle \Psi \vert \frac{d\hat{Q}(\theta_{j})}{d\theta_{j}}\vert \Phi \rangle)  $
    \State $\vert \Psi \rangle \leftarrow \hat{Q}(\theta_{j})^{-1} \vert \Psi \rangle $
\EndFor
\State Return $grads$

\end{algorithmic}
\end{algorithm}

We note that in steps 3 and 5 of Algorithm.\ref{alg:noiseless ad}, we have purposely written $Q(\theta_{j})^{-1}$ instead of $Q(\theta_{j})^{\dagger}$ (the latter is used in Ref.~\cite{GuoShang2023}). In the context of noiseless quantum circuits, these two expressions are equivalent, but the inverse expression used here can also be used for noisy quantum circuits where the quantum channel is in general not unitary.



\subsection{Simulating Noisy Quantum Circuits}\label{sec:noisy}

To describe the effects of noises, density matrices are required instead of pure states, and quantum channels should be used for the evolution of density matrices.
Nevertheless, for classical simulation purpose, we can utilize the same idea as the state vector simulator, by vectorizing these density matrices into state vectors. To be concrete, a density matrix $\rho$ for an n-qubit quantum system can be denoted as:
\begin{align}
  \rhoop=\sum_{\sigma_{n:1},\sigma_{n:1}^{\prime}}\rho_{\sigma_1,...,\sigma_n}^{\sigma_1^{\prime},...,\sigma_n^{\prime}}|\sigma_1,\ldots,\sigma_n)\langle\sigma_1^{\prime},\ldots,\sigma_n^{\prime}|.
\end{align}
where $\rho_{\sigma_1,...,\sigma_n}^{\sigma_1^{\prime},...,\sigma_n^{\prime}}$ forms a $2^n \times 2^n$ coefficient matrix. The vectorization operation transforms $\rhoop$ into an effective ``pure state'' (which is not normalized in general) as:
\begin{align}
|\rho\rangle\rangle = \sum_{\sigma_{n:1},\sigma_{n:1}^{\prime}} \tilde{\rho}_{\sigma_{1},\ldots,\sigma_{n},\sigma_{1}^{\prime},\ldots,\sigma_{n}^{\prime}} |\sigma_{1},\ldots,\sigma_{n},\sigma_{1}^{\prime},\ldots,\sigma_{n}^{\prime}\rangle.
\end{align}
Numerically, this vectorization is simply a reinterpretation of the data of the matrix $\rho_{\sigma_1,...,\sigma_n}^{\sigma_1^{\prime},...,\sigma_n^{\prime}}$ as a vector $\tilde{\rho}_{\sigma_{1},\ldots,\sigma_{n},\sigma_{1}^{\prime},\ldots,\sigma_{n}^{\prime}}$, no actual manipulation of the data is needed.
Correspondingly, the action of a quantum channel on a single qubit can be expressed as:
\begin{align}\label{eq:channel2}
\tilde{\Lambda}(\tilde{\rho}) = M_{\sigma_{j},\sigma_{j}^{\prime}}^{\tau_{j},\tau_{j}^{\prime}} \tilde{\rho}_{\sigma_{1},...,\sigma_{n},\sigma_{1}^{\prime},...,\sigma_{n}^{\prime}},
\end{align}
where $M_{\sigma_j,\sigma_j^{\prime}}^{\tau_j,\tau_j^{\prime}}=\sum_s K_{\tau_j^{\prime},\sigma_j^{\prime}}^s(K_{\tau_j,\sigma_j}^s)^*$ is a matrix derived from the Kraus operators as defined in Eq.(\ref{eq:channel}). 
We can see that Eq.(\ref{eq:channel2}) is analogous to a two-qubit quantum gate in Eq.(\ref{eq:twoqubitgate}) on a $2n$ qubit pure state, with the only difference that the operation $M_{\sigma_j,\sigma_j^{\prime}}^{\tau_j,\tau_j^{\prime}}$ may not be unitary.
Similarly, an $m$-qubit quantum channel can be mapped to a $2m$-qubit quantum gate operation on $|\rho\rangle\rangle$. In JuliVQC, this mapping is used extensively to simulate noisy quantum circuits.

\subsection{Noisy Automatic Differentiation}
Similar to the idea we used to simulate noisy quantum circuits, we will implement the backpropagation of noisy variational quantum circuits similar to the AD of noiseless quantum circuits by viewing an $n$-qubit density matrix $\rhoop$ as a $2n$-qubit ``pure'' quantum state $\vert \rho\rangle\rangle$.

First, the loss function for noisy quantum circuits can be generally written as 
\begin{align}\label{eq:loss_mixed}
\mathcal{L}(\theta) = \langle\langle I |\hat{H}\circuit(\theta)|0^{\otimes 2n}\rangle \rangle,
\end{align}
where $ \hat{I} $ denotes the identity matrix and $\vert I\rangle\rangle$ is the purification of it, and we have equivalently converted the final trace operation into the ``overlap'' operation with state $\vert I\rangle\rangle$. Each unitary gate operation $\hat{Q}_j$ in $\circuit(\theta)$ should be understood as quantum channel in Eq.(\ref{eq:channel}) with a single Kraus operator $K_1 = \hat{Q}_j$.
Eq.(\ref{eq:loss_mixed}) has a close correspondence to Eq.(\ref{eq:loss_pure}) for noiseless quantum circuits. Similar to Eq.(\ref{eq:diff_pure}), the derivative of Eq.(\ref{eq:loss_mixed}) against a specific parameter $\theta_j$ can be written as
\begin{align}\label{eq:diff_mixed}
  \frac{\partial\mathcal{L}(\theta)}{\partial\theta_j}=\langle\langle I|\hat{H}\circuit_{M:j+1}\frac{\mathrm{d}\hat{Q}(\theta_j)}{\mathrm{d}\theta_j}\circuit_{j-1:1}|0^{\otimes 2n}\rangle\rangle.
\end{align}
Drawing the connection between Eq.(\ref{eq:diff_pure}) and Eq.(\ref{eq:diff_mixed}),
we can still use a similar approach to Algorithm.\ref{alg:noiseless ad} to efficiently compute the gradient, by redefining the two intermediate states as: $\vert \Phi_j\rangle\rangle = \circuit_{j-1:1}|0^{\otimes 2n}\rangle$ and $\langle\langle \Psi_j\vert = \langle\langle I|\hat{H}\circuit_{M:j+1}$. This redefinition allows Eq.(\ref{eq:diff_mixed}) to be simplified into almost the same form as Eq.(\ref{eq:diff_pure2}):
\begin{align}\label{eq:diff_mixed2}
  \frac{\partial\mathcal{L}(\theta)}{\partial\theta_j}= \langle\langle \Psi_j\vert \frac{\mathrm{d}\hat{Q}(\theta_j)}{\mathrm{d}\theta_j} \vert \Phi_j\rangle\rangle.
\end{align}


\section{Performance}
\label{sec:Performance} 

This section presents a performance evaluation of JuliVQC in comparison with other leading quantum simulators: \texttt{Pennylane-lightning}, \texttt{Yao}, \texttt{Qiskit}, \texttt{ProjectQ}, and \texttt{Qulacs}. The benchmarks validate the algorithmic implementations discussed in Sec.\ref{sec:algs} and offer a comparative analysis of computational efficiency. For transparency and reproducibility, full benchmark results have been available in the associated repository~\cite{benchmark}.

\subsection{Experimental Setup}

  \begin{table}[]
    \centering
    \begin{tabular}{cc}
      \hline
        \textbf{ Library} & \textbf{Version} \\ \hline
        ProjectQ & 0.8.0 \\ 
        Qulacs & 0.6.3 \\ 
        Qiskit & 0.45.3 \\ 
        Qiskit-aer & 0.13.2 \\ 
        Pennylane & 0.34.0 \\ 
        Pennylane-lightning & 0.34.0 \\ 
        Yao & 0.8.13 \\
        JuliVQC & 0.0.1 \\\midrule
        Python & 3.9.18 \\ 
        Julia & 1.9.4 \\ 
        pytest-benchmark & 4.0.0 \\
        mkl & 2024.0.0 \\ 
        numpy & 1.26.3 \\ \hline
    \end{tabular}
  \caption{Libraries and their versions used in our benchmarks.}
  \label{tab1}
\end{table}

\begin{figure}[t]
  \centerline{\includegraphics[width=0.50\textwidth]{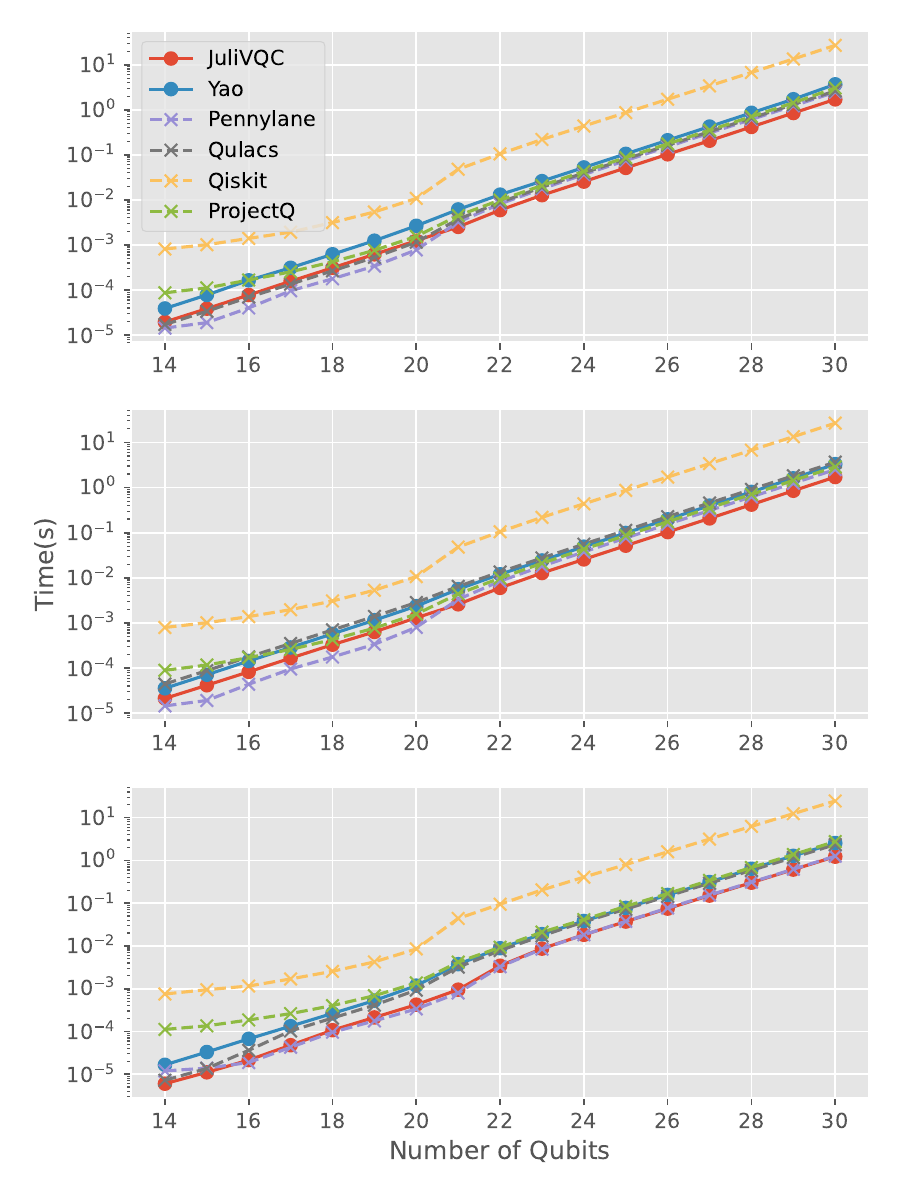}}
      \caption{A comparison of single-thread runtime performance for H (top), Rx (middle), and CNOT (bottom) gates.}
    \label{fig:single gate test}
  \end{figure}

\begin{figure*}[t]
  \centerline{\includegraphics[width=1\textwidth]{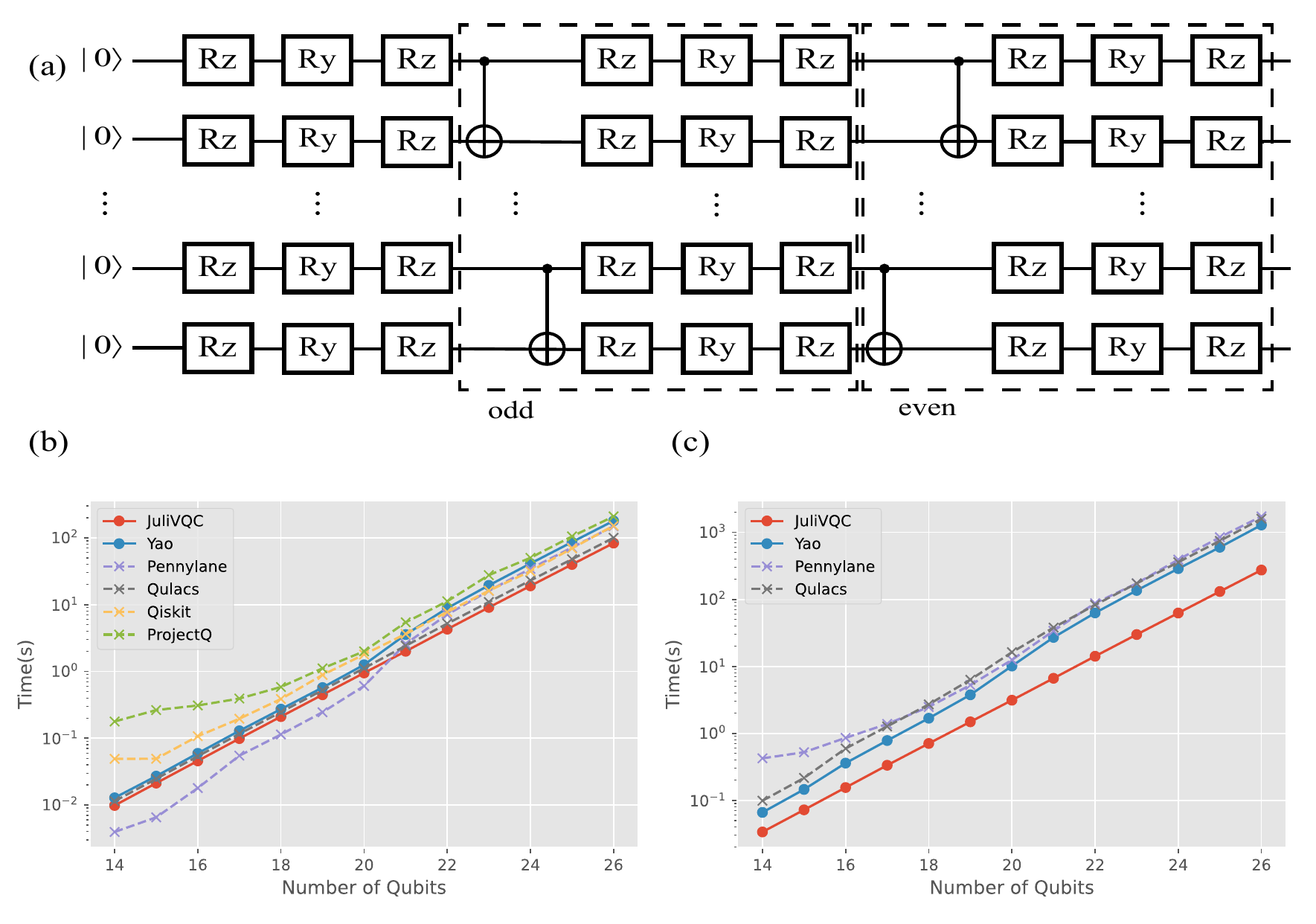}}
  \caption{
      (a) Structure of the random quantum circuit used for benchmarking, the gate operations inside the dashed box is counted as one layer.
      Runtime scaling of (b) simulating random quantum circuits and (c) computing gradients of random quantum circuits against the number of qubits.
  }
\label{fig:vqc_test}
\end{figure*}

We conducted the benchmarks on a workstation equipped with 4 Intel(R) Xeon(R) Gold 6254 CPUs, running Ubuntu 23.04.  The versions of the libraries and dependencies used are documented in Table.\ref{tab1}. Each simulator was updated to its latest release as of February 2024. .

For benchmarking, we employed the \texttt{pytest-benchmark} package for simulators with Python interfaces, specifically \texttt{PennyLane-lightning}, \texttt{Qulacs}, \texttt{Qiskit}, and \texttt{ProjectQ}. For simulators with Julia interfaces, namely \texttt{JuliVQC} and \texttt{Yao}, we used the \texttt{BenchmarkTools.jl} package.


\subsubsection{System Configuration}
To minimize fluctuations induced by the Linux operating system during benchmarks, we implemented the following measures:

\textbf{Processor Shielding}. We used \texttt{cset}, a Python wrapper for Linux cpuset pseudo-filesystem, to ensure processor shielding. By pinning processes to specific cpusets, this approach ensured that each trial was executed with a consistent processor/memory configuration.

\textbf{Virtual Memory Settings}. We optimized the virtual memory system for benchmarking by adjusting two key parameters. We set \texttt{vm.swappiness} to 10 to reduce swapping and disabled address space layout randomization (ASLR) by setting \texttt{kernel.randomize\underline{ }va\underline{ }space} to 0, ensuring consistent memory address assignments.

\textbf{CPU Frequency Scaling and Boosting}. To prevent potential skewing of benchmark results, we locked the CPU clock rate by setting all processors to the \texttt{performance} governor and disabled CPU boosting. This stabilized performance metrics and enhanced benchmark reproducibility.

\subsubsection{Python Call Overhead and Optimizations}

In our benchmarks, the function calls for simulators other than JuliVQC and Yao are made through Python interfaces, though the underlying implementations are written in C++. The overhead of invoking C++ functions from Python is a consideration that cannot be overlooked. Such overheads are typically in the order of 0.1 $\mu s$, which, while seemingly negligible, can be significant for small-scale quantum circuits. However, the smallest circuit in our benchmarks consists of 14 qubits with an execution time of the order of 10 $\mu s$. Consequently, we have chosen to disregard this additional overhead in our analysis.

We also utilized circuit optimization features provided by the simulators to minimize execution times. For example, Qulacs includes a \texttt{QuantumCircuitOptimizer()} function. We activated these optimizations during the benchmark tests to ensure peak simulator performance. It is important to note that the time spent on these optimizations was not included in the circuit simulation time calculations.

Despite those efforts to ensure fairness and consistency in benchmarking, we repeated each experiment 10 times and recorded the mean execution time as the final benchmark result, to minimize the fluctuation in a single trial.

\subsection{Single Gate Operation Performance}

We evaluated the basic performance of JuliVQC by testing its execution of three elementary quantum gates: the Hadamard (H), Pauli-X rotation (Rx), and Controlled NOT (CNOT) for quantum states ranging from 14 to 30 qubits. To ensure a precise assessment of individual gate operation efficiency, we disabled parallel processing features and conducted all tests on a single thread.
As expected, the execution time for the circuits increased exponentially with the number of qubits, aligning with the computational complexity typical of quantum simulations. Notably, JuliVQC demonstrated superior performance in executing H and Rx gates for circuits exceeding 21 qubits. For the CNOT gate, JuliVQC has a performance comparable to the leading simulator, \texttt{Pennylane-lightning} at all the tested scales.



\subsection{Quantum Circuit and Variational Quantum Circuit Performances}
\label{subsection:JuliVQC performance}


Quantum circuits and variational quantum circuits are essential ingredients for standard quantum algorithms and variational quantum algorithms respectively.
Here we test the efficiency of JuliVQC on these two scenarios by simulating random quantum circuits with structure in Fig.~\ref{fig:vqc_test}(a) and computing gradients for these circuits. 
For RQCs without computing gradients, we benchmark JuliVQC against \texttt{Yao}, \texttt{PennyLane-lightning}, \texttt{Qulacs}, \texttt{Qiskit}, \texttt{ProjectQ}.
For variational RQCs, 
we restrict our benchmarks to simulators with built-in AD support, specifically \texttt{Yao}, \texttt{PennyLane-lightning}, and \texttt{Qulacs}. Moreover, we used the Heisenberg Hamiltonian as the qubit operator for the latter benchmarks. The results for these two sets of benchmarks are depicted in Fig.~\ref{fig:vqc_test}(b,c) respectively. All benchmarks were conducted in a single-threaded environment.

The results indicate that JuliVQC excels in both simulating RQCs and computing their gradients. Remarkably, JuliVQC outperforms other simulators in simulating RQCs with over 22 qubits. It's superior efficiency is more evident for gradient calculations, which is attributed to the efficient implementation of the AD as detailed in Sec.~\ref{sec:algs}.
In both sets of benchmarks, we can see that JuliVQC is more advantageous for larger-scale quantum circuits.



\subsection{Parallelization Performance}

\begin{figure}[t]
  \centering
  \includegraphics[width=1.0
  \linewidth]{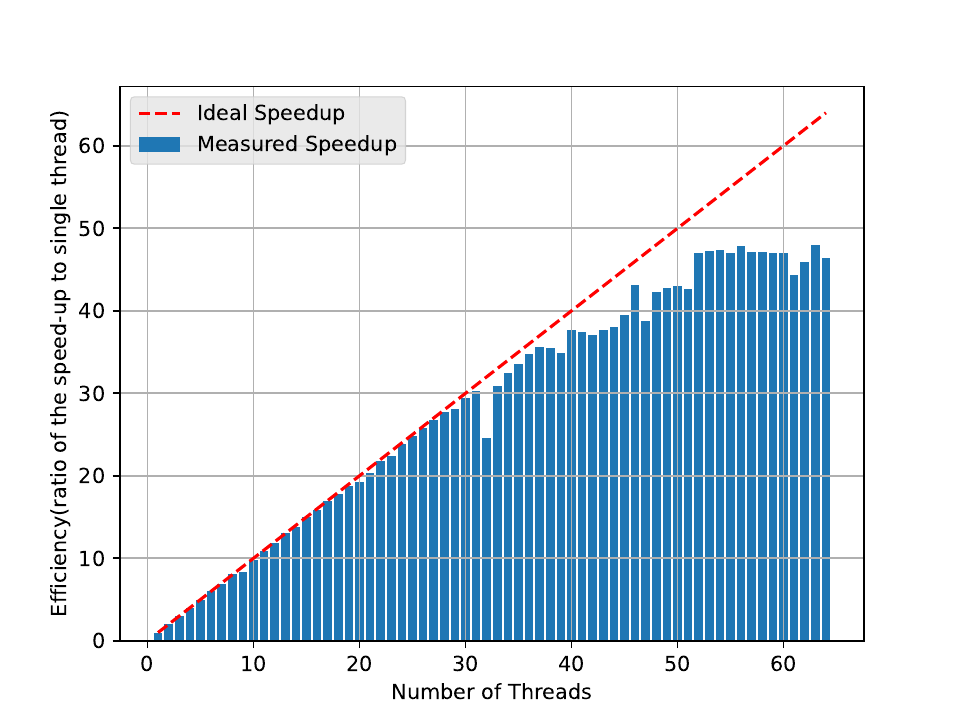}
  \caption{Parallelization performance of JuliVQC. The y-axis represents the acceleration ratio of multi-threaded JuliVQC compared to the single-threaded version.}
  \label{fig:parallelization_performance}
\end{figure}

In this subsection, we examine the parallelization capabilities of JuliVQC. The focus of this analysis is to benchmark the performance of JuliVQC for an increasing number of threads.
We utilized a 26-qubit RQC as shown in Fig.~\ref{fig:vqc_test}(a), to evaluate the performance of JuliVQC when using 1 to 64 threads. 
The results are illustrated in Fig.\ref{fig:parallelization_performance}, where each bar is averaged over $10$ simulations.
We can see that JuliVQC can achieve almost a perfect linear scaling with respect to the number of threads for less than $30$ threads, while beyond $30$ qubits the scaling becomes slower than linear and reach a plateau at around $50$ threads. 



\subsection{Noisy Automatic Differentiation}

\begin{figure}[t]
  \centering
  \includegraphics[width=1.0
  \linewidth]{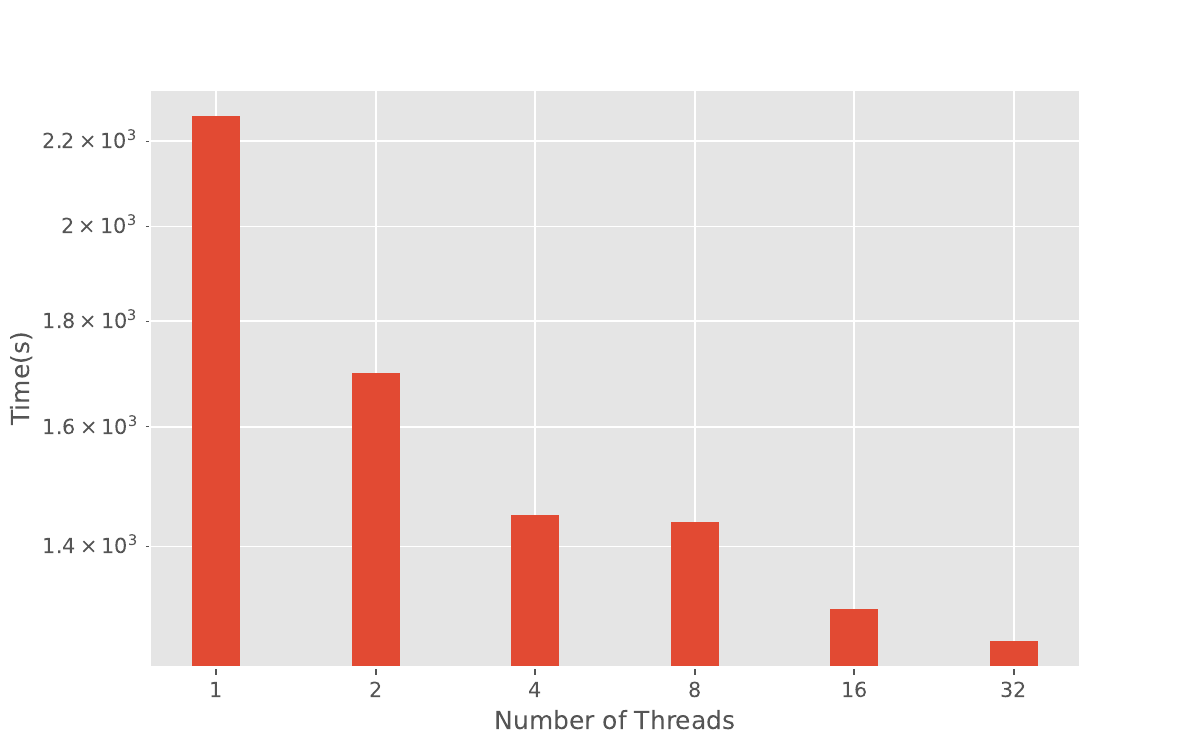}
  \caption{Results of noisy automatic differentiation test.}
  \label{fig:ad_performance}
\end{figure}

Finally, we test the performance of JuliVQC for simulating noisy quantum circuits. 
From Sec.~\ref{sec:noisy}, the efficiency of JuliVQC for an $n$-qubit noisy quantum circuit will be similar to that for an $2n$-qubit noiseless quantum circuit, as they use essentially the same implementation for each elementary gate operations. Therefore
here we only focus on testing the performance of JuliVQC for simulating the noisy variational quantum circuits. Since we are not aware of other simulators with built-in support for AD of noisy variational quantum circuits, we will only benchmark the multi-threading efficiency of JuliVQC against its single-threaded version, similar to Fig.~\ref{fig:parallelization_performance} . We use the same quantum circuits as in Fig.~\ref{fig:vqc_test}(a) with $14$ qubits, but we add a depolarizing channel on each qubit after each layer.

As shown in Fig.~\ref{fig:ad_performance}, the multi-threaded JuliVQC demonstrates fairly good speedup compared to the single-threaded version, similar to the noiseless case. However, the multi-threading performance for noise AD seems to the less stable compared to the noiseless case, as can be seen from the results with $8$ threads. The unstability of the multi-threading performance for noise AD could be due to that one needs to perform the matrix inverse in Algorithm.~\ref{alg:noiseless ad} (in the noiseless case this reduces to matrix adjoint for which no actual calculation needs to be done). With less than $8$ threads, the parallelization performance is more stable, as the amount of calculation distributed on each thread is larger and the difference between matrix inverse and matrix adjoint becomes negligible.


\section{Summary}\label{sec:summary}

In summary, we have presented an open-source quantum circuit simulator JuliVQC, which is light-weight but efficient, and seamlessly supports automatic differentiation for both noiseless and noise quantum circuits. We have shown the elementary procedures to run standard quantum algorithms and variational quantum algorithms using JuliVQC, as well as the algorithmic designs in JuliVQC which are he origin of its efficiency. JuliVQC can be used for researchers to test quantum algorithms and to develop noisy quantum computers, it can also be easily used as a backend for existing end-to-end quantum circuit simulation frameworks.

\bibliography{refs}

\end{document}